# Codebook Design Method for Noise Robust Speaker Identification based on Genetic Algorithm

Md. Rabiul Islam[1]

[1]Department of Computer Science & Engineering
Rajshahi University of Engineering & Technology
Rajshahi-6204, Bangladesh.
rabiul_cse@yahoo.com

Md. Fayzur Rahman[2]

[2]Department of Electrical & Electronic Engineering
Rajshahi University of Engineering & Technology
Rajshahi-6204, Bangladesh.
mfrahman3@yahoo.com

*Abstract—* **In this paper, a novel method of designing a codebook for noise robust speaker identification purpose utilizing Genetic Algorithm has been proposed. Wiener filter has been used to remove the background noises from the source speech utterances. Speech features have been extracted using standard speech parameterization method such as LPC, LPCC, RCC, MFCC, ΔMFCC and ΔΔMFCC. For each of these techniques, the performance of the proposed system has been compared. In this codebook design method, Genetic Algorithm has the capability of getting global optimal result and hence improves the quality of the codebook. Comparing with the NOIZEOUS speech database, the experimental result shows that 79.62 [%] accuracy has been achieved.**

*Keywords- Codebook Design; Noise Robust Speaker Identification; Genetic Algorithm; Speech Pre-processing; Speech Parameterization.*

## I. INTRODUCTION

Speaker Identification is the task of finding the identity of an unknown speaker among a stored database of speakers. There are various techniques to resolve the automatic speaker identification problem [1, 2, 3]. HMM is one of the most successful classifier for speaker identification system [4, 5]. To implement the speaker identification system in real time environment, codebook design is essential. The LBG algorithm is most popular to design the codebook due to its simplicity [6]. But the limitations of the LBG algorithm are the local optimal problem and its low speed. It is slow because for each iteration, determination of each cluster requires that each input vector be compared with all the codewords in the codebook. There were another methods such as modified K-means (MKM) algorithm [7], designing codewords from the trained vectors of each phoneme and grouping them together into a single codebook [8] etc. In codebook design, the above methods perform well in noiseless environments but the system performance degrades under noisy environments.

This paper deals the efficient approach for implementing the codebook design method for HMM based real time close-set text-dependent speaker identification system under noisy environments. To remove the background noise from the speech, wiener filter has been used. Efficient speech pre-processing techniques and different feature extraction techniques have been considered to improve the performance of this proposed noise robust codebook design method for speaker identification.

## II. SYSTEM OVERVIEW

The proposed codebook design method can be divided into two operations. One is the encoder and another is the decoder. The encoder takes the input speech utterance and outputs the index of the codeword considering the minimum distortion. To find out the minimum distortion, different types of genetic algorithm operations have been used. In decoding phase, when the decoder receives the index then it translates the index to its associate speaker utterance. Fig. 1 shows the block diagram of this proposed codebook design method.

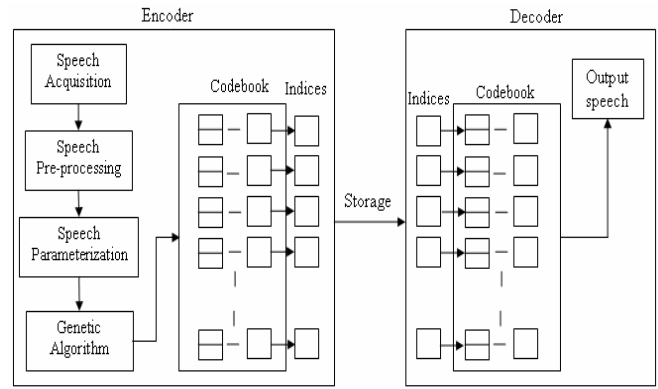

Figure 1. Paradigm of the proposed codebook design method.

## III. SPEECH SIGNAL PRE-PROCESSING

To capture the speech signal, sampling frequency of 11025 $H_Z$, sampling resolution of 16-bits, mono recording channel and Recorded file format = *.wav have been considered. The speech preprocessing part has a vital role for the efficiency of learning. After acquisition of speech utterances, winner filter has been used to remove the background noise from the original speech utterances [9, 10, 11]. Speech end points detection and silence part removal algorithm has been used to detect the presence of speech and to remove pulse and silences in a background noise [12, 13, 14, 15, 16]. To detect word boundary, the frame energy is computed using the sort-term log energy equation [17],

$$E_i = 10 \, \log \sum_{t=n_i}^{n_i + N - 1} S^2(t) \qquad (1)$$

Pre-emphasis has been used to balance the spectrum of voiced sounds that have a steep roll-off in the high frequency region [18, 19, 20]. The transfer function of the FIR filter in the z-domain is [19],

$$H(Z) = 1 - \alpha \cdot z^{-1}, \, 0 \le \alpha \le 1 \qquad (2)$$

Where $\alpha$ is the pre-emphasis parameter.

Frame blocking has been performed with an overlapping of 25% to 75% of the frame size. Typically a frame length of 10-30 milliseconds has been used. The purpose of the overlapping analysis is that each speech sound of the input sequence would be approximately centered at some frame [21, 22].

From different types of windowing techniques, Hamming window has been used for this system. The purpose of using windowing is to reduce the effect of the spectral artifacts that results from the framing process [23, 24, 25]. The hamming window can be defined as follows [26]:

$$w(n) = \begin{cases} 0.54 - 0.46 \cos \dfrac{2\Pi n}{N}, & -\left(\dfrac{N-1}{2}\right) \le n \le \left(\dfrac{N-1}{2}\right) \\ 0, & \text{Otherwise} \end{cases} \qquad (3)$$

## IV. SPEECH PARAMETERIZATION

This stage is very important in an ASIS because the quality of the speaker modeling and pattern matching strongly depends on the quality of the feature extraction methods. For the proposed ASIS, different types of speech feature extraction methods [27, 28, 29, 30, 31, 32] such as RCC, MFCC, ΔMFCC, ΔΔMFCC, LPC, LPCC have been applied.

## V. SPEECH PARAMETERIZATION

Genetic Algorithm [33, 34, 35, 36] has been applied in two ways for the encoding and decoding purposes. On encoding, every speaker utterance is compared with an environmental noise utterance and made some groups. In each group, one utterance is selected which is defined as the codeword of that group. As a result of encoding, some groups have been defined and one speaker utterance will lead one group. On decoding side, when unknown speaker utterance comes to the system then it is matched with a leading utterance. The unknown utterance will then find out within that selected group.

In GA processing selection, crossover and mutation operators have been used here. The fitness function is expressed as follows:

$$Fitness = (Unknown \, speech \times Each \, stored \, speech) \qquad (4)$$

In the recognition phase, for each unknown group and speaker within the group to be recognized, the processing shown in Fig. 2 has been carried out.

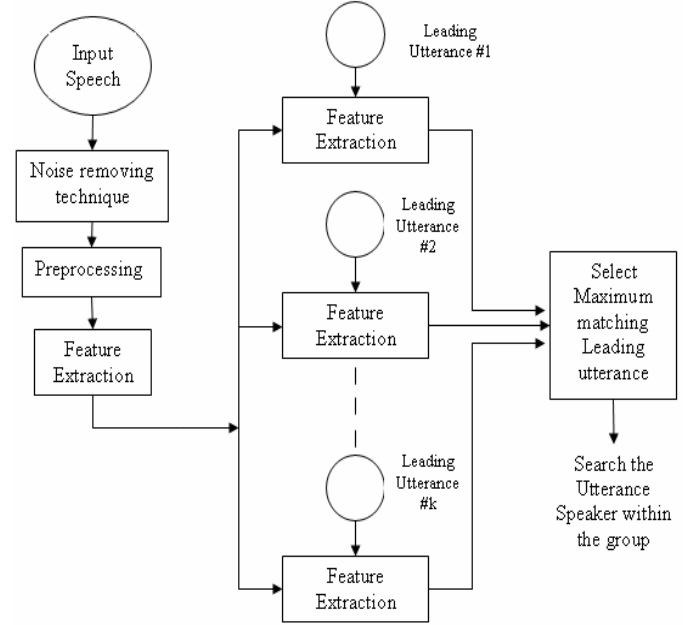

Figure 2.  Recognition model on Genetic Algorithm.

## VI. OPTIMUM PARAMETER SELECTION ON GENETIC ALGORITHM

### A. Experiment on the Crossover Rate

The identification rate has been measured according to the various crossover rates. Fig. 3 shows the comparison among results of different crossover rates. It is shown that the highest identification rate of 87.00 [%] was achieved at crossover rate 5.

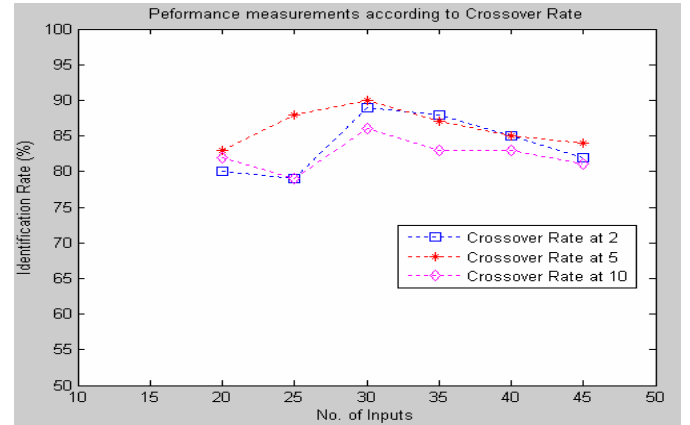

Figure 3.  Performance comparison among different crossover rate.

### B. Experiment on the No. of Generations

The number of generations has also been varied to measure the best performance of this codebook design method. According to the number of generation 5, 10 and 20 (with crossover rate 5), a comparative identification rate was found which is shown in Fig. 4. When the comparison is continued up to 5<sup>th</sup> generation, highest speaker identification rate of 93.00 [%] was achieved.

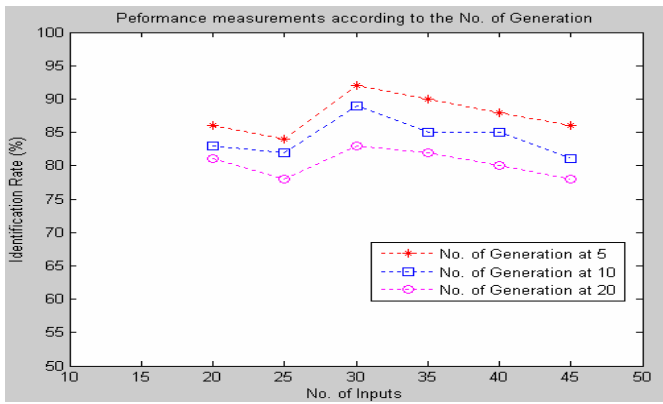

Figure 4. Performance comparison among various numbers of generations.

## VII. PERFORMANCE ANALYSIS OF THE PROPOSED CODEBOOK DESIGN METHOD

The optimal values of the critical parameters of the GA are chosen carefully according to various experiments. In noiseless environment, the crossover rate and number of generation have been found to be 5 for both. The performance analysis has been counted according to the text-dependent speaker identification system.

To measure the performance of the proposed system, NOIZEOUS speech database [37, 38] has been used. In NOIZEOUS speech database, eight different types of environmental noises (i.e. Airport, Babble, Car, Exhibition Hall, Restaurant, Street, Train and Train station) have been considered with four different SNRs such as 0dB, 5dB, 10dB and 15dB. All of the environmental conditions and SNRs have been accounted on the following experimental analysis.

TABLE I. AIRPORT NOISE AVERAGE IDENTIFICATION RATE (%) FOR NOIZEOUS SPEECH CORPUS

| Method SNR | MFCC | ΔMFCC | ΔΔMFCC | RCC | LPCC |
|---|---|---|---|---|---|
| 15dB | 89.00 | 86.33 | 63.33 | 65.33 | 75.67 |
| 10dB | 86.00 | 84.43 | 58.43 | 60.43 | 69.67 |
| 5dB | 75.33 | 81.00 | 50.33 | 60.33 | 60.43 |
| 0dB | 68.89 | 75.29 | 43.33 | 56.17 | 58.29 |
| Average | 79.81 | 81.76 | 53.86 | 60.57 | 65.93 |

TABLE II. BABBLE NOISE AVERAGE IDENTIFICATION RATE (%) FOR NOIZEOUS SPEECH CORPUS

| Method SNR | MFCC | ΔMFCC | ΔΔMFCC | RCC | LPCC |
|---|---|---|---|---|---|
| 15dB | 80.00 | 90.00 | 63.33 | 63.33 | 76.67 |
| 10dB | 76.67 | 86.67 | 53.33 | 56.67 | 70.00 |
| 5dB | 63.33 | 73.33 | 46.67 | 56.67 | 70.00 |
| 0dB | 73.33 | 63.33 | 46.67 | 53.33 | 63.33 |
| Average | 73.33 | 78.33 | 52.50 | 57.50 | 70.00 |

TABLE III. CAR NOISE AVERAGE IDENTIFICATION RATE (%) FOR NOIZEOUS SPEECH CORPUS

| Method SNR | MFCC | ΔMFCC | ΔΔMFCC | RCC | LPCC |
|---|---|---|---|---|---|
| 15dB | 76.67 | 89.43 | 63.33 | 73.33 | 76.67 |
| 10dB | 73.33 | 83.67 | 53.33 | 63.33 | 70.00 |
| 5dB | 63.33 | 73.33 | 53.33 | 63.33 | 70.00 |
| 0dB | 63.33 | 63.33 | 46.67 | 53.33 | 60.00 |
| Average | 69.17 | 77.44 | 54.17 | 63.33 | 69.17 |

TABLE IV. EXHIBITION HALL NOISE AVERAGE IDENTIFICATION RATE (%) FOR NOIZEOUS SPEECH CORPUS

| Method SNR | MFCC | ΔMFCC | ΔΔMFCC | RCC | LPCC |
|---|---|---|---|---|---|
| 15dB | 90.00 | 91.67 | 76.67 | 80.00 | 87.67 |
| 10dB | 83.33 | 83.33 | 63.33 | 76.67 | 76.67 |
| 5dB | 76.67 | 80.00 | 76.67 | 66.67 | 73.33 |
| 0dB | 73.33 | 76.67 | 53.33 | 63.33 | 70.00 |
| Average | 80.83 | 82.92 | 67.50 | 74.17 | 76.92 |

TABLE V. RESTAURANT NOISE AVERAGE IDENTIFICATION RATE (%) FOR NOIZEOUS SPEECH CORPUS

| Method SNR | MFCC | ΔMFCC | ΔΔMFCC | RCC | LPCC |
|---|---|---|---|---|---|
| 15dB | 85.00 | 91.00 | 53.33 | 83.33 | 83.33 |
| 10dB | 80.00 | 80.00 | 53.33 | 76.67 | 73.33 |
| 5dB | 73.33 | 76.67 | 50.43 | 63.33 | 73.33 |
| 0dB | 60.00 | 65.33 | 46.67 | 63.33 | 63.33 |
| Average | 74.58 | 78.25 | 50.94 | 71.67 | 73.33 |

TABLE VI. STREET NOISE AVERAGE IDENTIFICATION RATE (%) FOR NOIZEOUS SPEECH CORPUS

| Method SNR | MFCC | ΔMFCC | ΔΔMFCC | RCC | LPCC |
|---|---|---|---|---|---|
| 15dB | 83.33 | 90.00 | 63.33 | 76.67 | 83.33 |
| 10dB | 76.67 | 80.00 | 56.67 | 63.33 | 73.33 |
| 5dB | 73.33 | 76.67 | 53.33 | 76.67 | 73.33 |
| 0dB | 63.33 | 73.33 | 46.67 | 63.33 | 63.33 |
| Average | 74.17 | 80.00 | 55.00 | 70.00 | 73.33 |

TABLE VII. TRAIN NOISE AVERAGE IDENTIFICATION RATE (%) FOR NOIZEOUS SPEECH CORPUS

| Method SNR | MFCC | ΔMFCC | ΔΔMFCC | RCC | LPCC |
|---|---|---|---|---|---|
| 15dB | 90.00 | 91.33 | 63.33 | 73.33 | 85.00 |
| 10dB | 80.00 | 85.00 | 53.33 | 70.00 | 76.67 |
| 5dB | 66.67 | 86.67 | 53.33 | 63.33 | 63.33 |
| 0dB | 66.67 | 73.33 | 46.67 | 66.67 | 63.33 |
| Average | 75.84 | 84.08 | 54.17 | 68.33 | 72.08 |

TABLE VIII.    TRAIN STATION NOISE AVERAGE IDENTIFICATION RATE (%) FOR NOIZEOUS SPEECH CORPUS

| Method / SNR | MFCC | ΔMFCC | ΔΔMFCC | RCC | LPCC |
|---|---|---|---|---|---|
| 15dB | 86.67 | 90.00 | 53.33 | 70.00 | 76.67 |
| 10dB | 76.67 | 76.67 | 53.33 | 66.67 | 73.33 |
| 5dB | 63.33 | 66.67 | 46.67 | 56.67 | 63.33 |
| 0dB | 60.00 | 63.33 | 46.67 | 53.33 | 60.00 |
| Average | 71.67 | 74.17 | 50.00 | 61.67 | 68.33 |

Table IX shows the overall average speaker identification rate for NOIZEOUS speech corpus. From the table it is easy to compare the performance among MFCC, ΔMFCC, ΔΔMFCC, RCC and LPCC methods for DHMM based codebook technique. It is shown that ΔMFCC has greater performance (i.e. 79.62 [%]) than any other methods such as MFCC, ΔΔMFCC, RCC and LPCC.

TABLE IX.    OVERALL AVERAGE SPEAKER IDENTIFICATION RATE (%) FOR NOIZEOUS SPEECH CORPUS

| Method / Various Noises | MFCC | Δ MFCC | ΔΔ MFCC | RCC | LPCC |
|---|---|---|---|---|---|
| Airport Noise | 79.81 | 81.76 | 53.86 | 60.57 | 65.93 |
| Babble Noise | 73.33 | 78.33 | 52.50 | 57.50 | 70.00 |
| Car Noise | 69.17 | 77.44 | 54.17 | 63.33 | 69.17 |
| Exhibition Hall Noise | 80.83 | 82.92 | 67.50 | 74.17 | 76.92 |
| Restaurant Noise | 74.58 | 78.25 | 50.94 | 71.67 | 73.33 |
| Street Noise | 74.17 | 80.00 | 55.00 | 70.00 | 73.33 |
| Train Noise | 75.84 | 84.08 | 54.17 | 68.33 | 72.08 |
| Train Station Noise | 71.67 | 74.17 | 50.00 | 61.67 | 68.33 |
| Average Identification Rate (%) | 74.93 | 79.62 | 54.77 | 65.91 | 71.14 |

## VIII.    CONCLUSION AND OBSERVATION

The experimental results reveal that the performance of the proposed codebook design method yields about 93.00 [%] identification rate in noiseless environments and 79.62 [%] in noisy environments that are seemingly higher than the previous techniques that utilized LBG clustering method. However, a benchmark comparison is needed to establish the superiority of this proposed method and which is underway. In the speaker identification technique, noise is a common factor that influences the performance of this technique significantly. In this work, efficient noise removing technique has been used to enhance the performance of the proposed GA based codebook design method. So, GA based codebook design method is capable of protect in the system from noise distortion. The performance of this system may be tested by using large speech database and it will be the further work of this system.

AUTHORS PROFILE

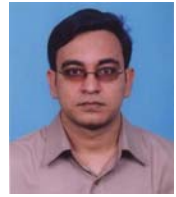

**Md. Rabiul Islam** was born in Rajshahi, Bangladesh, on December 26, 1981. He received his B.Sc. degree in Computer Science & Engineering and M.Sc. degrees in Electrical & Electronic Engineering in 2004, 2008, respectively from the Rajshahi University of Engineering & Technology, Bangladesh. From 2005 to 2008, he was a Lecturer in the Department of Computer Science & Engineering at Rajshahi University of Engineering & Technology. Since 2008, he has been an Assistant Professor in the Computer Science & Engineering Department, University of Rajshahi University of Engineering & Technology, Bangladesh. His research interests include bio-informatics, human-computer interaction, speaker identification and authentication under the neutral and noisy environments.

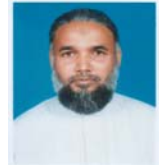

**Md. Fayzur Rahman** was born in 1960 in Thakurgaon, Bangladesh. He received the B. Sc. Engineering degree in Electrical & Electronic Engineering from Rajshahi Engineering College, Bangladesh in 1984 and M. Tech degree in Industrial Electronics from S. J. College of Engineering, Mysore, India in 1992. He received the Ph. D. degree in energy and environment electromagnetic from Yeungnam University, South Korea, in 2000. Following his graduation he joined again in his previous job in BIT Rajshahi. He is a Professor in Electrical & Electronic Engineering in Rajshahi University of Engineering & Technology (RUET). His current research interest are Dgital Sgnal Pocessing, Electronics & Machine Control and Hgh Vltage Dscharge Aplications. He is a member of the Institution of Engineer's (IEB), Bangladesh, Korean Institute of Illuminating and Installation Engineers (KIIEE), and Korean Institute of Electrical Engineers (KIEE), Korea.